\documentclass[11pt]{article}

\usepackage{acl}
\usepackage{fix-cm}
\usepackage{times}
\usepackage{latexsym}
\usepackage{todonotes}

\usepackage[T1]{fontenc}

\usepackage[utf8]{inputenc}

\usepackage{microtype}

\usepackage{inconsolata}

\usepackage{graphicx}

\title{CORVUS: Red-Teaming Hallucination Detectors via Internal Signal Camouflage in Large Language Models}

\usepackage{amsmath} 
\usepackage{amssymb}
\usepackage{booktabs}
\usepackage{multirow} 
\usepackage{algorithm}
\usepackage{algpseudocode}

\author{
  \textbf{Nay Myat Min\textsuperscript{1}},
  \textbf{Long H. Pham\textsuperscript{1}},
  \textbf{Hongyu Zhang\textsuperscript{2},
  \textbf{Jun Sun\textsuperscript{1}}
  }
  \\
  \\
  \textsuperscript{1}Singapore Management University \\
  \textsuperscript{2}Chongqing University \\
  \\
  \small{
    \textbf{Correspondence:} \href{mailto:myatmin.nay.2022@phdcs.smu.edu.sg}{myatmin.nay.2022@phdcs.smu.edu.sg}
  }
}

\begin{document}
\maketitle
\begin{abstract}
Single-pass hallucination detectors rely on internal telemetry (e.g., uncertainty, hidden-state geometry, and attention) of large language models, implicitly assuming hallucinations leave separable traces in these signals. We study a white-box, model-side adversary that fine-tunes lightweight LoRA adapters on the model while keeping the detector fixed, and introduce \textsc{CORVUS}, an efficient red-teaming procedure that learns to camouflage detector-visible telemetry under teacher forcing, including an embedding-space FGSM attention stress test. Trained on 1{,}000 out-of-distribution Alpaca instructions ($<0.5\%$ trainable parameters), \textsc{CORVUS} transfers to FAVA-Annotation across Llama-2, Vicuna, Llama-3, and Qwen2.5, and degrades both training-free detectors (e.g., LLM-Check) and probe-based detectors (e.g., SEP, ICR-probe), motivating adversary-aware auditing that incorporates external grounding or cross-model evidence.
\end{abstract}

\section{Introduction}
\label{sec:introduction}

Large language models (LLMs) remain prone to hallucination—fluent but unfounded content that impairs reliability in QA, summarization, and open-ended generation~\citep{10.1145/3571730,zhang2023siren,10.1145/3703155}. 
Broadly, hallucination defenses can be grouped into two categories: \emph{external knowledge--based} approaches, which query external tools or corpora (e.g., retrieval) and judge an answer by its agreement with retrieved evidence, and \emph{internal telemetry--based} approaches, which decide using only the model’s own computations. In this work, we focus on the latter class. 

Within internal telemetry--based approaches, a growing line of work studies \emph{single-pass} detectors: given a prompt and a realized answer, they use one teacher-forced forward pass to inspect token-level uncertainty and internal activations (such as hidden states, attentions). These methods implicitly assume hallucinations induce \emph{separable telemetry} that can be read out from one pass. We consider both training-free scores (e.g., PPL~\citep{renOutofDistributionDetectionSelective2022}, Window-Entropy~\citep{malininUncertaintyEstimationAutoregressive2020}, LLM-Check~\citep{sriramananLLMCheckInvestigatingDetection2024a}) and probe-based detectors (e.g., SEP~\citep{kossen2024semantic}, ICR-probe~\citep{zhang2025icrprobetrackinghidden}).

Prior evaluations typically treat the LLM as \emph{passive}: detectors may adapt to the model, but the model does not adapt in return. In open-weight or self-hosted deployments, a \emph{white-box, model-side} adversary can modify the LLM (e.g., lightweight adapter finetuning) to \emph{mask} detector-visible telemetry while keeping the surface answer unchanged. We therefore compare clean vs.\ adapted models under a \emph{fixed-answer, teacher-forced replay} protocol (\S\ref{sec:evaluation}), so detector differences reflect internal changes rather than different generations.

We introduce \emph{CORVUS} (\underline{C}amouflaging \underline{O}pen-weight \underline{R}epresentations, \underline{V}olumes, \underline{U}ncertainty, and \underline{S}tructure), a model-side red-teaming procedure that learns lightweight adapters to \emph{camouflage the internal telemetry} used by single-pass detectors. CORVUS directly optimizes three general telemetry signals over answer tokens: 
\emph{(i) hidden log-volume (HV)}, which summarizes how the answer-token representations spread out in hidden space; 
\emph{(ii) attention diagonality (AD)}, which measures how self-focused attention is within the answer span; and 
\emph{(iii) token entropy (TE)}, which captures how peaked the next-token distribution is. CORVUS uses a single-step Fast Gradient Sign Method (FGSM)~\citep{goodfellow2015explainingharnessingadversarialexamples} in embedding space to construct a telemetry-directed \emph{stress test} for attention diagonality: the perturbation is crafted to \emph{increase} an attention-based detector score on the embeddings.

\textbf{Contributions.} We introduce \textsc{CORVUS} and evaluate single-pass hallucination detectors under a white-box, model-side adversary that efficiently adapts the model via lightweight adapters.

\begin{itemize}
  \item \textbf{Method.} We define answer-window telemetry (TE/HV/AD) and train LoRA adapters under teacher forcing to reshape it, including a one-step embedding-space FGSM attention term.

  \item \textbf{OOD training.} We train \textsc{CORVUS} only on 1{,}000 out-of-distribution Alpaca instructions and never use FAVA-Annotation~\citep{mishra2024finegrained} prompts for training; the same adapters transfer to FAVA without further tuning.

  \item \textbf{Evasion results.} On FAVA-Annotation across four open-weight LLMs (Llama-2, Vicuna, Llama-3, Qwen2.5), \textsc{CORVUS} degrades all the existing detectors calibrated on clean telemetry, including training-free scores (LLM-Check~\citep{sriramananLLMCheckInvestigatingDetection2024a}) and trained probes (ICR-probe~\citep{zhang2025icrprobetrackinghidden}, SEP~\citep{kossen2024semantic}).
  
  \item \textbf{Mechanism.} We connect controlled shifts in TE/HV/AD to failures of fixed detector scores, showing how modest adapter updates can collapse telemetry separability (\S\ref{sec:mechanistic}).
\end{itemize}

Intuitively, single-pass telemetry-based detectors resemble exit polls: they infer outcomes from process traces rather than auditing external evidence. \textsc{CORVUS} is the strategic respondent---it adapts the model so that, for the same surface answer, detector-visible telemetry looks typical and becomes less distinguishable to a fixed detector.

\section{Background: Hallucination Detection}

\paragraph{Training-free defenses.}
Sampling-based approaches like SelfCheckGPT~\citep{manakul-etal-2023-selfcheckgpt} assess consistency across multiple generations but incur substantial latency. Single-pass uncertainty metrics such as per-token perplexity~\citep{renOutofDistributionDetectionSelective2022} and token-entropy-based scores~\citep{malininUncertaintyEstimationAutoregressive2020} are efficient but can be brittle to calibration and prompt shifts. Representation-centric methods summarize structure in attentions and hidden states (e.g., eigenvalue or log-determinant surrogates~\citep{Chen2024INSIDELI}); LLM-Check~\citep{sriramananLLMCheckInvestigatingDetection2024a} combines uncertainty with log-determinant features from hidden-state covariance (\emph{Hidden Score}) and attention kernels (\emph{Attention Score}) under teacher forcing.

\paragraph{Training-based defenses.}
Probe methods train lightweight classifiers over hidden states~\citep{azaria-mitchell-2023-internal,ch-wang-etal-2024-androids}. Semantic Entropy Probes (SEP)~\citep{kossen2024semantic} fit a small head on a selected hidden state to predict semantic-entropy-derived targets, enabling single-pass inference without sampling, but can be sensitive to dataset shift. ICR-probe~\citep{zhang2025icrprobetrackinghidden} summarizes cross-layer residual-stream dynamics by comparing attention-guided attribution to actual hidden updates, then pools layerwise discrepancies with a small MLP.

Across these detectors, a shared premise holds: hallucinations leave \emph{distinguishable telemetry} in the model’s computations, including uncertainty shifts, changes in hidden-state geometry, and attention structure (and, for ICR-probe, residual--attention mismatch). In our notation, these correspond to TE/HV/AD and related cross-layer signals. This shared premise motivates a model-side adversary that learns to camouflage such telemetry without materially altering the generated text. 

\paragraph{Retrieval-augmented defenses.} External knowledge based approaches, alignment scoring with external references (e.g., AlignScore~\citep{zha2023alignscore}) and RAG-specific mechanisms (e.g., ReDeEP~\citep{sun2025redeep}) can be strong when high-quality evidence exists, but they change the threat surface and may not apply to open-ended generation or settings where retrieval is unavailable. 

\section{The Proposed CORVUS Method}
\label{sec:methods}

\subsection{Method overview}
\label{sec:method-overview}
\begin{figure*}[!t]
\centering
\includegraphics[width=0.92\textwidth]{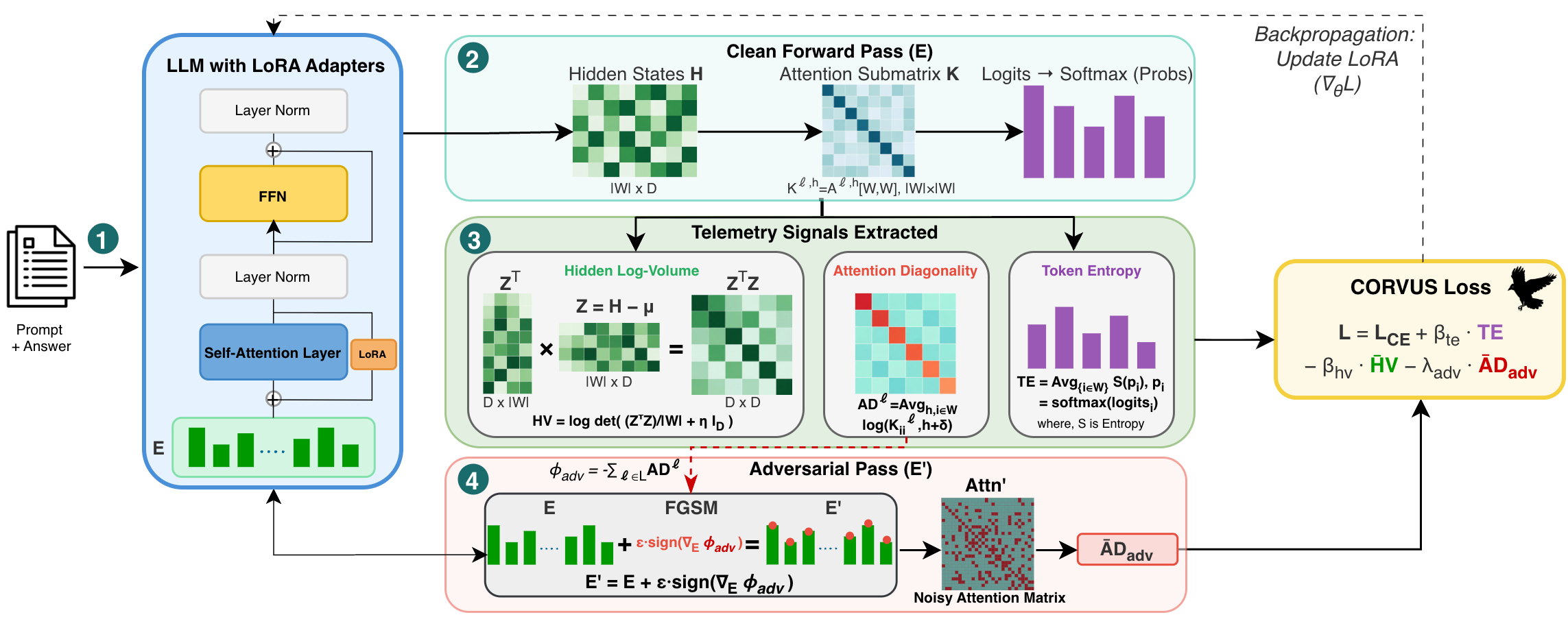}
\caption{\textbf{CORVUS method.} \textbf{(1)} A prompt and fixed answer are processed under teacher forcing with trainable LoRA adapters. \textbf{(2)} A clean pass yields logits, hidden states, and attention over the answer window. \textbf{(3)} Telemetry features are extracted: TE, HV, and AD. \textbf{(4)} An AD-targeted single-step FGSM embedding perturbation induces a perturbed pass to obtain $AD_{\mathrm{adv}}$; a combined loss updates only LoRA to reshape detector-relevant internal signals.}
\label{fig:corvus_figure}
\end{figure*}

CORVUS is a model-side red-teaming procedure that trains lightweight adapters to camouflage the telemetry used by single-pass hallucination detectors. With the detector fixed and the base model frozen, CORVUS optimizes only the adapters under the auditor’s teacher-forced replay on fixed (prompt, answer) pairs: a clean pass computes TE/HV/AD over the answer window, an AD-targeted FGSM embedding perturbation induces a stress-test pass, and a joint loss preserves teacher-forced likelihood while shifting telemetry. As the objective is label-free, adapters can be trained on an out-of-distribution dataset and still transfer to the audited dataset (see Figure~\ref{fig:corvus_figure} and Algorithm~\ref{alg:CORVUS}).

\subsection{Threat model}
\label{sec:threat-model}
We consider an adaptive, model-side adversary (red team) with white-box access to hidden states and attentions and the ability to fine-tune lightweight adapters. The goal is to reduce the hallucination confidence of a telemetry-based fixed detector under single-pass, teacher-forced auditing while preserving task utility and, to the extent possible, the user-visible answer content. We focus on post-hoc auditing where an auditor evaluates a \emph{realized} answer string via teacher-forced replay; correspondingly, our primary evaluation holds the answer string fixed when comparing clean vs.\ CORVUS-adapted models. This isolates internal-signal camouflage from textual changes. We realistically assume that the adversary has no hallucination-labeled examples from the target benchmark distribution, so CORVUS trains only on an out-of-distribution instruction set and optimizes telemetry terms that transfer to the auditing setting. We target a practical budget (single GPU) and do not rely on multiple generations or external retrieval. 

\subsection{Terminology}
\label{sec:terminology}

We consider decoder-only LLMs evaluated under \emph{teacher forcing}. For an input sequence of $T$ tokens, the model produces layerwise hidden states $\{X^{\ell}\}_{\ell=0}^{L}$ and attention probabilities $\{A^{\ell,h}\}_{\ell=1,h=1}^{L,H}$. Here $X^{\ell} \in \mathbb{R}^{T \times D}$ denotes hidden states at layer $\ell$, where $D$ is the model hidden size (with $X^{0}$ denoting input embeddings), and $A^{\ell,h} \in \mathbb{R}^{T \times T}$ denotes the head-wise attention distributions (after softmax) for head $h$ at layer $\ell$. When a single attention matrix is needed, we use the head-averaged attention $\mathrm{Attn}^{\ell} = \frac{1}{H}\sum_{h=1}^{H} A^{\ell,h}$.

We focus on answer tokens in an \emph{answer window} $W \subseteq \{0,\dots,T-1\}$. During CORVUS training on instruction--response pairs, the label mask identifies answer positions and we take $W$ to be the set of non-ignore indices. For evaluation, we tokenize the prompt $x_p$ (length $p$) and the answer string $x$ to form the concatenated sequence $x_p \oplus x$ with total length $T$, and set $W \,=\, \{p, p+1, \dots, T-1\},$
i.e., all tokens corresponding to the answer.

\subsection{Telemetry over internal signals}
\label{sec:telemetry}

We compute three general differentiable telemetry signals over $W$ in one teacher-forced pass: token entropy (TE), hidden log-volume (HV), and attention diagonality (AD). TE captures token uncertainty, HV summarizes answer-token dispersion, and AD measures self-focused attention within the answer span; together they form a compact summary exploited by many detectors and CORVUS.

\paragraph{Token entropy (TE).}
Let $L \in \mathbb{R}^{T\times V}$ be teacher-forced logits over a vocabulary of size $V$, and let $p_i = \mathrm{softmax}(L_i)$ denote the predictive distribution at token position $i$. We define \emph{Token Entropy (TE)} as the mean token-level predictive entropy over the answer window $W$:
{
\begin{equation}
\mathrm{TE} \,=\, \frac{1}{|W|}\sum_{i\in W} \Big(\,- \sum_{v=1}^{V} p_i[v]\,\log p_i[v]\,\Big).
\end{equation}
}
TE is the mean categorical entropy of the next-token distribution over answer tokens (often known as logit entropy); windowed variants take a maximum to emphasize localized uncertainty spikes.

\paragraph{Hidden log-volume (HV).}
For each layer $\ell$, let $X^{\ell}_W \in \mathbb{R}^{|W|\times D}$ be the slice of $X^{\ell}$ over answer tokens. We first center hidden states across tokens, $Z = X^{\ell}_W - \mathbf{1}\mu^\top$ with $\mu = \frac{1}{|W|}\sum_{i\in W} X^{\ell}_i \in \mathbb{R}^{D}$ (each $X^{\ell}_i \in \mathbb{R}^{D}$). We then form a feature-space covariance proxy
{
\begin{equation}
C_{\mathrm{hid}}^{\ell} \,=\, \frac{Z^\top Z}{|W|} + \eta I_D,\quad \eta>0,
\end{equation}
}
and define the \emph{HV} as the log-determinant
{
\begin{equation}
\mathrm{HV}^{\ell} \,=\, \log\det\!\big(C_{\mathrm{hid}}^{\ell}\big).
\end{equation}
}
We use a small jitter ($\eta=10^{-3}$) for numerical stability and to handle the common case $|W|<D$ (rank-deficient covariance without regularization). This logdet-style proxy is the hidden-state regularizer used in our CORVUS implementation and is closely related to log-volume surrogates used by training-free detectors such as LLM-Check.

\paragraph{Attention diagonality (AD).}
For each layer $\ell$ and head $h$, let $K^{\ell,h} = A^{\ell,h}[W, W] \in \mathbb{R}^{|W|\times |W|}$ denote the attention submatrix restricted to answer tokens.
We summarize diagonal concentration by
{
\small
\begin{equation}
\mathrm{AD}^{\ell} \,=\, \frac{1}{H}\sum_{h=1}^{H}\Bigg[\,\frac{1}{|W|}\sum_{i\in W} \log\big(K^{\ell,h}_{ii}+\delta\big)\,\Bigg],\quad \delta>0.
\end{equation}
}
We use $\delta = 10^{-8}$ for numerical stability to avoid $\log 0$ from (near-)zero diagonal attention probabilities. Higher $\mathrm{AD}^{\ell}$ corresponds to more self-diagonal attention mass, and averaging across layers yields a scalar closely related to the diagonal component of LLM-Check's Attention Score.

\subsection{Detector assumptions and signals}
\label{sec:detector-assumptions}
We briefly summarize the primary signals used by representative single-pass detectors and how they relate to TE/HV/AD:

\begin{itemize}
  \item \textbf{PPL}~\citep{renOutofDistributionDetectionSelective2022}: a sequence-level proxy of output uncertainty derived from per-token negative log-likelihood of the token, closely tied to TE over the answer window.
  \item \textbf{Window-Entropy}~\citep{malininUncertaintyEstimationAutoregressive2020}: an entropy-based uncertainty score over the answer window. We use a windowed token-entropy variant and take the maximum over windows, which reduces to the maximum token entropy over the answer window.
  \item \textbf{LLM-Check}~\citep{sriramananLLMCheckInvestigatingDetection2024a}: a training-free detector that aggregates output-uncertainty features with representation-centric features. In our notation, its Hidden Score is a global log-volume surrogate closely related to HV (a mean log-determinant / log-spectrum of a hidden-state covariance proxy), and its Attention Score corresponds to a global variant of AD (mean log of attention diagonal mass), both computed under teacher forcing.
  \item \textbf{SEP}~\citep{kossen2024semantic,kuhn2023semantic}: a probe trained to predict semantic-entropy-derived targets from a single hidden state, coupling TE-like uncertainty with representation information at a chosen layer.
  \item \textbf{ICR-probe}~\citep{zhang2025icrprobetrackinghidden}: a probe over \emph{cross-layer} residual-stream dynamics that compares attention-guided attribution to actual hidden-state updates and pools these discrepancies across layers with a small MLP.
\end{itemize}

\paragraph{Observation.}
On clean models evaluated on FAVA-Annotation, TE/HV/AD exhibit consistent distribution shifts between faithful and hallucinated answers, and these shifts are exploitable by single-pass detectors. CORVUS targets this vulnerability by inducing internal-signal shifts so that clean-calibrated decision rules (e.g., fixed thresholds) do not transfer, and in some cases the telemetry becomes less distinguishable. 

\subsection{Adversarial embedding red-teaming}
\label{sec:adv-embedding}

FGSM is used as a telemetry-directed surrogate: we craft a one-step embedding perturbation to \emph{increase} an attention-based detector score proportional to low attention diagonality (i.e., to \emph{decrease} $\mathrm{AD}$). This does not imply the perturbed pass induces semantic hallucinations or incorrect answers. Concretely, we apply a single-step perturbation in embedding space and then include a loss term that encourages higher $\mathrm{AD}$ on the perturbed pass. Let $E$ denote the input embeddings for a batch of tokenized sequences. We define the FGSM objective on a forward pass with differentiable embeddings as
{
\begin{equation}
\phi_{\mathrm{adv}} \,=\, -\sum_{\ell\in\mathcal{L}} \mathrm{AD}^{\ell},
\end{equation} 
}
and compute its gradient with respect to $E$. An FGSM update constructs perturbed embeddings
{
\begin{equation}
E' \,=\, E + \varepsilon\,\mathrm{sign}\big( \nabla_{E} \, \phi_{\mathrm{adv}} \big), \quad \varepsilon>0,
\end{equation} 
}
which \emph{increase} $\phi_{\mathrm{adv}}$ and thus \emph{decrease} attention diagonality. We then run a forward pass with $E'$ (keep model parameters shared) and compute attention telemetry $\{\mathrm{AD}^{\ell}_{\mathrm{adv}}\}$ under this perturbation.

\subsection{Training objective}
\label{sec:training-objective}
Figure~\ref{fig:corvus_figure} provides an overview of the CORVUS training procedure and Algorithm~\ref{alg:CORVUS} summarizes the full training loop. Let $\overline{(\cdot)}$ denote layerwise means on the clean or adversarial forward (e.g., $\overline{\mathrm{HV}}_{\mathrm{clean}} = \frac{1}{|\mathcal{L}|}\sum_{\ell\in\mathcal{L}}\mathrm{HV}^{\ell}$ over a set of monitored layers $\mathcal{L}$). For a given batch, the CORVUS loss is
{
\begin{equation}
\begin{aligned}
\mathcal{L}
&= \mathcal{L}_{\mathrm{CE}}
   + \beta_{\mathrm{te}}\,\mathrm{TE}_{\mathrm{clean}} \\
&\quad 
   - \beta_{\mathrm{hv}}\,\overline{\mathrm{HV}}_{\mathrm{clean}} - \lambda_{\mathrm{adv}}\,\overline{\mathrm{AD}}_{\mathrm{adv}}\,.
\end{aligned}
\label{eq:methods-obj}
\end{equation}
}where $\mathcal{L}_{\mathrm{CE}}$ is the standard sequence-level cross-entropy loss under teacher forcing, and $\beta_{\mathrm{te}}, \beta_{\mathrm{hv}}, \lambda_{\mathrm{adv}} > 0$ are scalar coefficients. The TE term penalizes token entropy on the clean pass, the HV term is maximized to adversarially shift hidden-geometry features, and the adversarial AD term $-\lambda_{\mathrm{adv}}\,\overline{\mathrm{AD}}_{\mathrm{adv}}$ encourages higher diagonality on the perturbed pass. The FGSM direction is computed by differentiating the surrogate detector score $-\overline{\mathrm{AD}}$ with respect to the input embeddings. All telemetry terms are computed under teacher forcing over the answer window, matching the detector feature-extraction protocol.

\begin{algorithm}[t]
\small
\caption{CORVUS training}
\label{alg:CORVUS}
\begin{algorithmic}[1]
\Require Model $M$; tokenizer $\mathcal{T}$; train data $\mathcal{D}$; steps $S_{\max}$
\Require LoRA modules $\mathcal{O}$ (rank $r$, scaling $\alpha$); penalties $(\beta_{\mathrm{te}},\beta_{\mathrm{hv}})$; FGSM step size $\varepsilon$; $\lambda_{\mathrm{adv}}$; telemetry options (e.g., layer set $\mathcal{L}$)
\State Attach LoRA($M, \mathcal{O}, r, \alpha$); freeze base parameters
\For{$s=1$ to $S_{\max}$}
  \State Sample $(\text{prompt}, \text{answer}) \sim \mathcal{D}$; tokenize with $\mathcal{T}$ to obtain \text{targets $y$}, window $W$, and embeddings $E$
  \State \textbf{Clean pass:} run $M$ under teacher forcing; collect logits, hidden states, and attentions
  \State $\mathrm{TE}_{\mathrm{clean}} \leftarrow \textsc{TokenEntropy}(W,\text{logits})$
  \State $(\overline{\mathrm{HV}}_{\mathrm{clean}}, \overline{\mathrm{AD}}_{\mathrm{clean}}) \leftarrow$
  \Statex $\quad\quad\quad\quad\quad\quad\quad\quad \textsc{Telemetry}(W,\text{hidden},\text{attn},\mathcal{L})$
  \State $\phi_{\mathrm{adv}} \leftarrow -\sum_{\ell\in\mathcal{L}}\mathrm{AD}^{\ell}$ \hfill\textit{// detector score proxy}
  \State $E' \leftarrow E + \varepsilon\,\mathrm{sign}(\nabla_E \, \phi_{\mathrm{adv}})$ \hfill\textit{// FGSM}
  \State \textbf{Adversarial pass:} run $M$ with $E'$; collect attn$'$
  \State $\overline{\mathrm{AD}}_{\mathrm{adv}} \leftarrow \textsc{Telemetry}(W,\cdot,\text{attn}',\mathcal{L})$ \hfill\textit{// AD only}
  \State $\mathcal{L}_{\mathrm{CE}} \leftarrow \textsc{CE\_Loss}(\text{logits}, \text{y})$
  \State $\mathcal{L} \leftarrow \mathcal{L}_{\mathrm{CE}} + \beta_{\mathrm{te}}\,\mathrm{TE}_{\mathrm{clean}}$ 
  \Statex $\quad\quad\quad\quad - \beta_{\mathrm{hv}}\,\overline{\mathrm{HV}}_{\mathrm{clean}} - \lambda_{\mathrm{adv}}\,\overline{\mathrm{AD}}_{\mathrm{adv}}$
  \State Backpropagate $\nabla_{\theta_{\mathrm{LoRA}}} \mathcal{L}$; update LoRA parameters
\EndFor
\end{algorithmic}
\end{algorithm}

\section{Training and Implementation Details}
\label{sec:impl}

\paragraph{Telemetry computation.}
We compute token-level \emph{TE} and per-layer \emph{HV} and \emph{AD} on the answer tokens (as defined by the label mask / answer window), following \S\ref{sec:telemetry}. We average HV/AD across a monitored layer set $\mathcal{L}$ on-the-fly. For the adversarial term, gradients flow through the adversarial pass (via $\overline{\mathrm{AD}}_{\mathrm{adv}}$) into LoRA parameters; clean telemetry values are used as loss terms but are not explicitly matched between passes.

\paragraph{Adapter fine-tuning.}
CORVUS uses lightweight adapter-based fine-tuning to preserve base model capacity while optimizing the evasion objective.
In our runs (implemented via LLaMA-Factory~\citep{zheng2024llamafactory}), we attach LoRA adapters with rank $r{=}64$ and scaling $\alpha{=}128$, and target only the attention projection modules $(q\_proj, k\_proj)$. All base model parameters are frozen.
Training uses bfloat16, gradient accumulation (4 steps), gradient checkpointing, and AdamW with learning rate $2\!\times\!10^{-4}$ and a cosine learning-rate scheduler.

\paragraph{Objective weighting.}
The loss uses penalties $(\beta_{\mathrm{te}}, \beta_{\mathrm{hv}})$ and $\lambda_{\mathrm{adv}}$ (Eq.~\ref{eq:methods-obj}). HV enters with a negative sign to increase hidden log-volume along the monitored layer band; TE enters with a positive sign to penalize high token entropy over response tokens; and the adversarial attention term enters with a negative sign on $\overline{\mathrm{AD}}_{\mathrm{adv}}$, which increases attention diagonality on the perturbed pass over the same layer band. Our default coefficients are: $\beta_{\mathrm{te}}=0.07$, $\beta_{\mathrm{hv}}=0.05$, and $\lambda_{\mathrm{adv}}=0.5$.

\paragraph{Adversarial step.}
We apply FGSM with magnitude $\varepsilon{=}1.0\!\times\!10^{-2}$ at each update. We first run an embedding-gradient pass to compute $\nabla_E \phi_{\mathrm{adv}}$, and then run a second pass with perturbed embeddings to compute the adversarial \emph{AD} term (Eq.~\ref{eq:methods-obj}). The perturbed embeddings are not clamped or projected back to the original embedding manifold.

\paragraph{Data and batching.}
We train CORVUS adapters on the Alpaca\_en dataset~\citep{alpaca} via LLaMA-Factory. This corpus is \emph{out-of-distribution} relative to FAVA-Annotation: it consists of generic instruction–response pairs without hallucination labels, and we do not use any FAVA-Annotation prompts or labels when optimizing the CORVUS objective. The base loss is standard sequence-level cross-entropy under teacher forcing; the label mask defines the answer tokens for TE/HV/AD.

\paragraph{Compute and measurement.}
Experiments run on a single GPU node (H100), 8 CPU cores, and 48\,GB system RAM. We train for a single epoch over the 1{,}000 Alpaca examples with cosine learning-rate scheduling. The resulting adapters are then evaluated zero-shot on FAVA-Annotation.
We next describe the detectors, benchmark, and fixed-answer replay protocol used to evaluate CORVUS.

\section{Evaluation Setup}
\label{sec:evaluation}

\subsection{Detectors and Baselines}
We evaluate CORVUS against two families of state-of-the-art single-pass detectors under a teacher-forced replay on fixed (prompt, answer) pairs:

\begin{itemize}
  \item \textbf{Training-free detectors.} We consider standard single-pass uncertainty and representation-centric scores: PPL, Window-Entropy (max token entropy over the answer window; $w{=}1$), and LLM-Check (Hidden/Attention Scores).
  
  \item \textbf{Training-based detectors.} We also consider SEP (single-state probe) and ICR-probe (MLP over layerwise residual--attention mismatch). Probes are trained on clean-model telemetry (FAVA-train) and tested on held-out examples.
\end{itemize}

\subsection{Benchmark and Models}
We ground our evaluation in the FAVA-Annotation dataset introduced by \citep{mishra2024finegrained}, which consists of prompts and model-generated answers annotated with sequence-level hallucination labels.
We follow the split protocol from LLM-Check for training, validation, and test partitions. We experiment with four open-weight, instruction-tuned LLMs commonly used in hallucination-detection work:
Llama-2-7B-Chat~\citep{touvron2023llama2openfoundation}, Vicuna-7B~\citep{zheng2023judging}, Llama-3-8B-Instruct~\citep{grattafiori2024llama3herdmodels}, and Qwen2.5-14B-Instruct~\citep{qwen2025qwen25technicalreport}. For each model, we associate each FAVA-Annotation prompt with a benchmark-provided answer string.

\subsection{Protocol}
For each model, we compare a clean base model (no adapters) to a CORVUS-adapted model (LoRA fine-tuned on Alpaca). All detectors are computed via teacher-forced replay on a given (prompt, answer) pair. We use a \emph{fixed-answer replay} protocol: for each prompt, we reuse the benchmark-provided answer string when extracting telemetry from both models. Importantly, it is also a conservative restriction for an attacker: allowing benign rephrasing or changing the answer distribution can only make evasion easier, whereas our goal is to show that detector reliability can fail due to internal-signal shifts even when the surface text is held constant. Training-based detectors (SEP, ICR-probe) are \emph{trained only} on clean-model telemetry from the FAVA-Annotation training split and then evaluated unchanged on both clean and CORVUS telemetry (auditor fixed; model changes). CORVUS adapters are trained only on Alpaca, so performance on FAVA reflects out-of-distribution transfer.

\paragraph{Evaluation metrics.}
For each detector and model on FAVA-Annotation, we report: \emph{AUROC}, as a threshold-independent measure of hallucination detection performance, \emph{Accuracy} and \emph{F1} at an operating point chosen on the validation set; and \emph{Fixed-threshold deltas}: calibrate a threshold on the clean model at a target false-positive rate (e.g., $5\%$ FPR) and evaluate the same threshold under CORVUS, reporting changes in accuracy, FPR, TPR, and AUROC. We do not require the CORVUS-adapted model to reproduce the same answer under free decoding; generation-time changes are orthogonal to the post-hoc auditing we target.

\paragraph{Utility preservation sanity check.}
Although CORVUS is trained to reshape telemetry under teacher forcing (and is not optimized to preserve free-decoding generations), we verify that the learned adapters do not induce a trivial generation failure mode under free decoding. In a representative run (500 prompts, greedy), both the base and CORVUS-adapted models produced \emph{non-empty} responses with \emph{no refusal-style outputs}, and we did not observe repetition collapse (95th-percentile maximum consecutive word-run of 2). This indicates that detector evasion is not explained by a degenerate ``non-generating'' or ``refusal-only'' collapse, and that the adapted model retains generation utility (readable, instruction-following outputs). As a lightweight surface-form fluency proxy, we also compute external-LM perplexity (GPT-2~\citep{radford2019language}) on the answers; the median perplexity shifts modestly from 11.1 (base) to 13.7 (ours). 

\section{Results}
\begin{table*}[t]
\centering
\small
\setlength{\tabcolsep}{6pt}
\begin{tabular}{@{}l c c c c@{}}
\toprule
\textbf{Metric} & \textbf{AUROC} & \textbf{Accuracy} & \textbf{TPR@5\%FPR} & \textbf{F1} \\
\midrule
\multicolumn{5}{c}{\textbf{Llama-2-7B-Chat}} \\
\midrule

PPL Score            & 53.18 $\rightarrow$ 48.09 (\textcolor{red}{-5.10})   & 58.68 $\rightarrow$ 50.90 (\textcolor{red}{-7.78}) & 3.59 $\rightarrow$ 0.00 (\textcolor{red}{-3.59})    & 68.92 $\rightarrow$ 5.75 (\textcolor{red}{-63.17}) \\
Window-Entropy       & 56.99 $\rightarrow$ 55.30 (\textcolor{red}{-1.69})   & 57.19 $\rightarrow$ 48.50 (\textcolor{red}{-8.69}) & 7.19 $\rightarrow$ 1.20 (\textcolor{red}{-5.99})    & 42.11 $\rightarrow$ 2.27 (\textcolor{red}{-39.84}) \\
Hidden (L17)   & 58.15 $\rightarrow$ 54.10 (\textcolor{red}{-4.04})  & 56.89 $\rightarrow$ 50.00 (\textcolor{red}{-6.89}) & 8.98 $\rightarrow$ 0.00 (\textcolor{red}{-8.98})   & 64.36 $\rightarrow$ 0.00 (\textcolor{red}{-64.36}) \\
Attn Score (L20)     & 73.63 $\rightarrow$ 48.08 (\textcolor{red}{-25.54})  & 69.76 $\rightarrow$ 50.00 (\textcolor{red}{-19.76}) & 10.18 $\rightarrow$ 0.00 (\textcolor{red}{-10.18})   & 69.85 $\rightarrow$ 0.00 (\textcolor{red}{-69.85}) \\
SEP            & 61.79 $\rightarrow$ 44.89 (\textcolor{red}{-16.90})  & 52.36 $\rightarrow$ 50.00 (\textcolor{red}{-2.36}) & 4.05 $\rightarrow$ 3.38 (\textcolor{red}{-0.67})   & 66.51 $\rightarrow$ 6.67 (\textcolor{red}{-59.84}) \\
ICR Probe            & 71.44 $\rightarrow$ 52.36 (\textcolor{red}{-19.08})  & 71.74 $\rightarrow$ 34.78 (\textcolor{red}{-36.96}) & 98.31 $\rightarrow$ 30.77 (\textcolor{red}{-67.54})   & 81.69 $\rightarrow$ 40.00 (\textcolor{red}{-41.69}) \\
\midrule
\multicolumn{5}{c}{\textbf{Vicuna-7B (v1.5)}} \\
\midrule
PPL Score            & 54.07 $\rightarrow$ 50.40 (\textcolor{red}{-3.67})  & 49.40 $\rightarrow$ 36.30 (\textcolor{red}{-13.10}) & 3.59 $\rightarrow$ 0.00 (\textcolor{red}{-3.59})   & 6.63 $\rightarrow$ 0.00 (\textcolor{red}{-6.63}) \\
Window-Entropy       & 45.43 $\rightarrow$ 47.10 (\textcolor{blue}{+1.67}) & 48.50 $\rightarrow$ 35.70 (\textcolor{red}{-12.80}) & 0.60 $\rightarrow$ 1.40 (\textcolor{blue}{+0.80})   & 1.15 $\rightarrow$ 2.63 (\textcolor{blue}{+1.48}) \\
Hidden (L13)   & 58.19 $\rightarrow$ 58.00 (\textcolor{red}{-0.19}) & 52.40 $\rightarrow$ 36.30 (\textcolor{red}{-16.10}) & 9.58 $\rightarrow$ 0.00 (\textcolor{red}{-9.58})  & 16.75 $\rightarrow$ 0.00 (\textcolor{red}{-16.75}) \\
Attn Score (L18)     & 71.91 $\rightarrow$ 59.70 (\textcolor{red}{-12.21}) & 60.78 $\rightarrow$ 36.30 (\textcolor{red}{-24.48}) & 25.75 $\rightarrow$ 0.00 (\textcolor{red}{-25.75})  & 39.63 $\rightarrow$ 0.00 (\textcolor{red}{-39.63}) \\
SEP            & 67.63 $\rightarrow$ 44.78 (\textcolor{red}{-22.85})  & 54.05 $\rightarrow$ 50.34 (\textcolor{red}{-3.72}) & 11.49 $\rightarrow$ 2.03 (\textcolor{red}{-9.46}) & 41.38 $\rightarrow$ 3.92 
(\textcolor{red}{-37.46}) \\
ICR Probe            & 69.70 $\rightarrow$ 62.00 (\textcolor{red}{-7.70}) & 67.39 $\rightarrow$ 35.90 (\textcolor{red}{-31.49}) & 18.50 $\rightarrow$ 0.10 (\textcolor{red}{-18.40})  & 79.45 $\rightarrow$ 36.20 (\textcolor{red}{-43.25}) \\
\midrule
\multicolumn{5}{c}{\textbf{Llama-3-8B-Instruct}} \\
\midrule
PPL Score            & 53.22 $\rightarrow$ 50.00 (\textcolor{red}{-3.22})  & 58.68 $\rightarrow$ 36.30 (\textcolor{red}{-22.38}) & 3.59 $\rightarrow$ 0.00 (\textcolor{red}{-3.59})   & 67.40 $\rightarrow$ 0.00 (\textcolor{red}{-67.40}) \\
Window-Entropy      & 60.17 $\rightarrow$ 62.10 (\textcolor{blue}{+1.93}) & 56.59 $\rightarrow$ 36.10 (\textcolor{red}{-20.49}) & 2.99 $\rightarrow$ 0.30 (\textcolor{red}{-2.69})   & 55.52 $\rightarrow$ 0.60 (\textcolor{red}{-54.92}) \\
Hidden (L21)   & 57.10 $\rightarrow$ 57.00 (\textcolor{red}{-0.01}) & 51.80 $\rightarrow$ 36.30 (\textcolor{red}{-15.50}) & 7.78 $\rightarrow$ 0.00 (\textcolor{red}{-7.78})  & 14.24 $\rightarrow$ 0.00 (\textcolor{red}{-14.24}) \\
Attn Score (L22)     & 68.95 $\rightarrow$ 58.50 (\textcolor{red}{-10.45}) & 53.89 $\rightarrow$ 36.30 (\textcolor{red}{-17.59}) & 12.57 $\rightarrow$ 0.00 (\textcolor{red}{-12.57})  & 21.89 $\rightarrow$ 0.00 (\textcolor{red}{-21.89}) \\
SEP            & 68.31 $\rightarrow$ 41.21 (\textcolor{red}{-27.10})  & 57.09 $\rightarrow$ 47.30 (\textcolor{red}{-9.80})  & 5.41 $\rightarrow$ 3.38 (\textcolor{red}{-2.03})     & 56.36 $\rightarrow$ 7.14 
(\textcolor{red}{-49.21}) \\
ICR Probe            & 72.78 $\rightarrow$ 53.00 (\textcolor{red}{-19.78})  & 69.57 $\rightarrow$ 35.00 (\textcolor{red}{-34.57}) & 25.75 $\rightarrow$ 8.00 (\textcolor{red}{-17.75})  & 79.10 $\rightarrow$ 39.00 (\textcolor{red}{-40.10}) \\
\midrule
\multicolumn{5}{c}{\textbf{Qwen2.5-14B-Instruct}} \\
\midrule
PPL Score            & 53.30 $\rightarrow$ 49.80 (\textcolor{red}{-3.50})  & 49.10 $\rightarrow$ 35.90 (\textcolor{red}{-13.20}) & 1.80 $\rightarrow$ 0.00 (\textcolor{red}{-1.80})   & 3.41 $\rightarrow$ 0.00 (\textcolor{red}{-3.41}) \\
Window-Entropy       & 60.87 $\rightarrow$ 60.20 (\textcolor{red}{-0.67}) & 52.40 $\rightarrow$ 41.20 (\textcolor{red}{-11.20}) & 8.98 $\rightarrow$ 0.00 (\textcolor{red}{-8.98})   & 15.87 $\rightarrow$ 0.00 (\textcolor{red}{-15.87}) \\
Hidden (L33)   & 60.44 $\rightarrow$ 58.88 (\textcolor{red}{-1.56}) & 51.50 $\rightarrow$ 36.30 (\textcolor{red}{-15.20}) & 7.19 $\rightarrow$ 0.00 (\textcolor{red}{-7.19})  & 12.90 $\rightarrow$ 0.00 (\textcolor{red}{-12.90}) \\
Attn Score (L37)     & 68.06 $\rightarrow$ 46.50 (\textcolor{red}{-21.56}) & 53.89 $\rightarrow$ 36.30 (\textcolor{red}{-17.59}) & 11.98 $\rightarrow$ 0.00 (\textcolor{red}{-11.98})  & 20.62 $\rightarrow$ 0.00 (\textcolor{red}{-20.62}) \\
SEP            & 68.90 $\rightarrow$ 46.20 (\textcolor{red}{-22.70})  & 56.80 $\rightarrow$ 35.90 (\textcolor{red}{-20.90}) & 9.00 $\rightarrow$ 0.00 (\textcolor{red}{-9.00})   & 58.50 $\rightarrow$ 0.00 (\textcolor{red}{-58.50}) \\
ICR Probe            & 77.14 $\rightarrow$ 60.00 (\textcolor{red}{-17.14})  & 66.80 $\rightarrow$ 36.30 (\textcolor{red}{-30.50}) & 16.00 $\rightarrow$ 0.00 (\textcolor{red}{-16.00})   & 70.00 $\rightarrow$ 42.16 (\textcolor{red}{-27.84}) \\
\bottomrule
\end{tabular}
\caption{\textbf{Results.} Values are Clean $\rightarrow$ After (\(\Delta\)). Positive deltas (blue) indicate improved detection; negative (red) indicate successful evasion. \textit{Hidden} and \textit{Attn Scores} denote the corresponding components from \textsc{LLM-Check}.}
\label{tab:main_results}
\end{table*}

We evaluate CORVUS on \emph{FAVA-Annotation} with four open-weight LLMs. Table~\ref{tab:main_results} compares \emph{clean} vs.\ \emph{CORVUS-adapted} models under identical prompts and fixed-answer teacher-forced replay. For each detector/model pair, we report AUROC, accuracy, TPR at a fixed FPR, F1, and the change induced by attaching CORVUS adapters. Across all four models on FAVA-Annotation, CORVUS reduces the effectiveness of both training-free and training-based detectors. Training-free scores (PPL, Window-Entropy, and LLM-Check and its components) and trained probes (SEP and ICR-probe) degrade when applied to CORVUS-adapted telemetry, despite probes being trained only on clean-model traces. We also report fixed-threshold deltas (clean-calibrated at 5\% FPR) for entropy, attention, and hidden components. Thresholds are chosen on clean-model validation telemetry and then held fixed under CORVUS. Clean-calibrated thresholds can yield much higher FPR and lower TPR after adaptation, indicating brittleness to model-side attacks. Across models, clean-calibrated operating points often collapse (TPR@5\%FPR near 0 and F1 near 0), revealing substantial calibration drift under model-side adaptation. Attention/representation-centric signals are the most brittle: the LLM-Check Attention Score loses up to 25.5 AUROC points (73.6$\rightarrow$48.1 on Llama-2), and SEP/ICR-probe drop by 17--27 and up to 20 AUROC points, respectively. 

\paragraph{Calibration shift vs.\ separability.}
To distinguish calibration shift from separability loss, we report both AUROC and fixed-threshold metrics. Large fixed-threshold drops with smaller AUROC changes suggest calibration shift (clean-calibrated thresholds no longer transfer), while AUROC drops indicate reduced separability. We observe both effects across telemetry families. Additional ablations are reported in the Appendix~\ref{sec:ablations}.

\begin{table*}[t]
\centering
\small
\setlength{\tabcolsep}{13pt}
\begin{tabular}{@{}p{0.10\textwidth}p{0.22\textwidth}p{0.55\textwidth}@{}}
\toprule
\textbf{Detector} & \textbf{Telemetry read} & \textbf{CORVUS-affected telemetry} \\
\midrule
PPL & Per-token log-likelihood / PPL over answer window & TE term directly alters the token-level predictive distribution (and hence entropy/perplexity) under teacher forcing on the answer window. \\
\hline
Window-Entropy & Windowed token entropy over the answer window (max over windows) & Driven by entropy: CORVUS changes the token-level predictive distribution under teacher forcing (hence entropy statistics), while shifts in HV/AD that change routing can indirectly modify these distributions. \\
\hline
LLM-Check & Hidden log-det (\emph{Hidden Score}); attention-kernel log-det (\emph{Attention Score}) & HV term reshapes hidden log-volume features that feed the Hidden Score; AD term shapes diagonal attention patterns (and thus the attention log-det component). \\
\hline
SEP & Single-state probe predicting semantic-entropy-derived uncertainty & TE term changes uncertainty-related behavior of the model; HV and AD terms change the hidden representation at the probed layer from which SEP reads out, affecting the features the probe relies on. \\
\hline
ICR-probe & Layerwise JSD between residual-induced projection distributions and attention & AD influences the structure and concentration of attention distributions in the JSD; HV changes the hidden directions along which residual updates are measured, altering the projection-induced distributions. \\
\bottomrule
\end{tabular}
\caption{\textbf{Telemetry map.} Summary of which telemetry features different detectors primarily rely on and which components of the CORVUS objective interact with those features. }
\label{tab:detector-mapping}
\end{table*}

\section{Mechanistic View of CORVUS}
\label{sec:mechanistic}

To interpret the degradations in Table~\ref{tab:main_results}, we provide a mechanistic view of why internal telemetry is, in general, not a reliable signal for hallucination detection under adaptive model-side attacks.

\paragraph{Detectors as functionals of telemetry.}
Let $T(x)\in\mathbb{R}^m$ denote single-pass, layer-aggregated telemetry features (e.g., token entropy, hidden-state geometry, attention structure, cross-layer residual--attention mismatch) computed on an input--output pair $x$, and let $s(x)=g(T(x))$ be the detector score. Training-free methods use a fixed aggregation $g(\cdot)$, while probe-based methods learn a shallow $g(\cdot)$ (e.g., MLP) on top of $T(x)$.
Detection succeeds when the between-class separation
$
\Delta \;\triangleq\; \big\|\; \mathbb{E}_{x\in\mathcal{H}}[T(x)] \;-\; \mathbb{E}_{x\in\mathcal{F}}[T(x)] \;\big\|
$
is large, where $\mathcal{H}$ and $\mathcal{F}$ denote hallucinated and faithful instances, respectively.
Empirically, on clean models and FAVA-Annotation, we observe such separability for TE, HV, and AD when used as part of LLM-Check-style features and probes.

\paragraph{How CORVUS changes telemetry.}
CORVUS modifies the model, not the detector: it fine-tunes lightweight adapters so that, under the same teacher-forced protocol, the induced telemetry $T(x)$ for hallucinated and faithful instances becomes less separable.
The loss directly interacts with three components of $T(x)$:
(i) the clean-pass token entropy over the answer window, $\mathrm{TE}_{\text{clean}}$;
(ii) a clean-pass hidden log-volume term, $\overline{\mathrm{HV}}_{\text{clean}}$, that summarizes answer-window hidden Gram matrices across selected layers; and
(iii) an adversarial attention-diagonal term, $\overline{\mathrm{AD}}_{\text{adv}}$, computed after a single-step FGSM perturbation in embedding space. By penalizing $\mathrm{TE}_{\text{clean}}$ and shaping $\overline{\mathrm{HV}}_{\text{clean}}$ and $\overline{\mathrm{AD}}_{\text{adv}}$, CORVUS shifts the telemetry distribution that detectors read—without modifying the detector architecture or its training data. Consequently, the effective separation $\Delta$ shrinks and the score $s(x)=g(T(x))$ becomes less informative about hallucination labels, even when the surface answer is held fixed. Table~\ref{tab:detector-mapping} summarizes which telemetry components different detectors primarily read and which parts of the CORVUS affect them.

\paragraph{Why this does not reduce to ``making the model dull.''}
A trivial evasion strategy would be to degrade the model (e.g., collapse activations or saturate logits), which harms task utility and is easily detectable. CORVUS avoids this by (i) retaining the standard teacher-forced cross-entropy loss, (ii) using small telemetry coefficients so LoRA updates remain low-rank and low-magnitude, and (iii) applying the adversarial term via a one-step embedding perturbation rather than enforcing extreme attention patterns on clean inputs. We report a utility preservation sanity check in \S\ref{sec:evaluation} to confirm the adapters do not trivially break generation.

\paragraph{Implications.}
Our results indicate that detectors should be evaluated not only on static, clean models but also under adaptive, model-side adversaries that can reshape internal telemetry. Robust defenses will likely require (i) diversity in telemetry signals that are harder to jointly manipulate with lightweight adapters, (ii) some form of randomization or challenge--response that makes white-box minimization more difficult, and (iii) cross-model validation (e.g., frozen auditors) to reduce the impact of on-model representation-level evasion.

\section{Conclusion}
\label{sec:conclusion}
We introduced \textsc{CORVUS}, a lightweight LoRA red-teaming method that camouflages single-pass telemetry (TE/HV/AD, with FGSM) under teacher-forced replay. Trained on 1{,}000 out-of-distribution Alpaca examples without detector labels, \textsc{CORVUS} transfers to FAVA-Annotation and degrades training-free scores (e.g., LLM-Check) and probe-based detectors (e.g., SEP, ICR-probe) across multiple open-weight models. Our results suggest that internal telemetry alone is not reliable for hallucination detection under adaptive model-side attacks, implying that external knowledge--based approaches, while challenging, may be inevitable for robust auditing. The open-source implementation of \textsc{CORVUS} is available at this~\href{https://figshare.com/s/327f6299606913e1b10e}{URL}.

\section*{Limitations}
\label{sec:limitations}
CORVUS assumes white-box access to the LLM and is most applicable to open-source or self-hosted models where an adversary can extract representations and fine-tune lightweight adapters; it does not directly apply to closed APIs, except via surrogate-model attacks. Our evaluation is limited to single-pass detectors operating on one model’s internals; more advanced defenses (e.g., retrieval-based fact-checking) were out of scope and may counter this attack. Finally, CORVUS targets the model’s internals, not its factuality: we do not optimize for correctness, and the attack can hide a hallucination without fixing it. Developing attacks that improve factuality, or defenses that explicitly monitor factual consistency, are left for future work.

\section*{Ethical Considerations}
This research is a form of adversarial red-teaming aimed at improving LLM safety. We demonstrate how an aligned LLM’s internal traces can be manipulated to evade hallucination detectors in a controlled setting. Our intent is to expose this vulnerability so that detector designers can patch it, not to facilitate misuse. We release our code in a responsible manner. All experiments were conducted on public datasets (Alpaca and FAVA-Annotation datasets) in accordance with their licenses. No private or user data was used. We strongly discourage any malicious use of CORVUS to bypass real-world safety systems. Instead, we hope this work informs the development of more robust hallucination defenses. For example, by incorporating adversarial training or cross-checks, ultimately leading to safer and more trustworthy AI systems.

\bibliography{custom}

@article{mishra2024finegrained,
      author = {Mishra, Abhika and Asai, Akari and Balachandran, Vidhisha and Wang, Yizhong and Neubig, Graham and Tsvetkov, Yulia and Hajishirzi, Hannaneh},
      title  = {Fine-grained Hallucinations Detections},
      journal = {arXiv preprint},
      year = {2024},
      url = {https://arxiv.org/abs/2401.06855}
    }

@misc{goodfellow2015explainingharnessingadversarialexamples,
      title={Explaining and Harnessing Adversarial Examples}, 
      author={Ian J. Goodfellow and Jonathon Shlens and Christian Szegedy},
      year={2015},
      eprint={1412.6572},
      archivePrefix={arXiv},
      primaryClass={stat.ML},
      url={https://arxiv.org/abs/1412.6572}, 
}

@inproceedings{azaria-mitchell-2023-internal,
    title = "The Internal State of an {LLM} Knows When It{'}s Lying",
    author = "Azaria, Amos  and
      Mitchell, Tom",
    editor = "Bouamor, Houda  and
      Pino, Juan  and
      Bali, Kalika",
    booktitle = "Findings of the Association for Computational Linguistics: EMNLP 2023",
    month = dec,
    year = "2023",
    address = "Singapore",
    publisher = "Association for Computational Linguistics",
    url = "https://aclanthology.org/2023.findings-emnlp.68/",
    doi = "10.18653/v1/2023.findings-emnlp.68",
    pages = "967--976"
}

@inproceedings{ch-wang-etal-2024-androids,
    title = "Do Androids Know They{'}re Only Dreaming of Electric Sheep?",
    author = "CH-Wang, Sky  and
      Van Durme, Benjamin  and
      Eisner, Jason  and
      Kedzie, Chris",
    editor = "Ku, Lun-Wei  and
      Martins, Andre  and
      Srikumar, Vivek",
    booktitle = "Findings of the Association for Computational Linguistics: ACL 2024",
    month = aug,
    year = "2024",
    address = "Bangkok, Thailand",
    publisher = "Association for Computational Linguistics",
    url = "https://aclanthology.org/2024.findings-acl.260/",
    doi = "10.18653/v1/2024.findings-acl.260",
    pages = "4401--4420"
}

@misc{
kossen2024semantic,
title={Semantic Entropy Probes: Robust and Cheap Hallucination Detection in {LLM}s},
author={Jannik Kossen and Jiatong Han and Muhammed Razzak and Lisa Schut and Shreshth A Malik and Yarin Gal},
year={2025},
url={https://openreview.net/forum?id=YQvvJjLWX0}
}

@inproceedings{
Chen2024INSIDELI,
title={{INSIDE}: {LLM}s' Internal States Retain the Power of Hallucination Detection},
author={Chao Chen and Kai Liu and Ze Chen and Yi Gu and Yue Wu and Mingyuan Tao and Zhihang Fu and Jieping Ye},
booktitle={The Twelfth International Conference on Learning Representations},
year={2024},
url={https://openreview.net/forum?id=Zj12nzlQbz}
}

@article{10.1145/3571730,
author = {Ji, Ziwei and Lee, Nayeon and Frieske, Rita and Yu, Tiezheng and Su, Dan and Xu, Yan and Ishii, Etsuko and Bang, Ye Jin and Madotto, Andrea and Fung, Pascale},
title = {Survey of Hallucination in Natural Language Generation},
year = {2023},
issue_date = {December 2023},
publisher = {Association for Computing Machinery},
address = {New York, NY, USA},
volume = {55},
number = {12},
issn = {0360-0300},
url = {https://doi.org/10.1145/3571730},
doi = {10.1145/3571730},
journal = {ACM Comput. Surv.},
month = mar,
articleno = {248},
numpages = {38}
}

@article{zhang2023siren,
    author = {Zhang, Yue and Li, Yafu and Cui, Leyang and Cai, Deng and Liu, Lemao and Fu, Tingchen and Huang, Xinting and Zhao, Enbo and Zhang, Yu and Chen, Yulong and Wang, Longyue and Luu, Anh Tuan and Bi, Wei and Shi, Freda and Shi, Shuming},
    title = {Siren’s Song in the AI Ocean: A Survey on Hallucination in Large Language Models},
    journal = {Computational Linguistics},
    pages = {1-46},
    year = {2025},
    month = {09},
    issn = {0891-2017},
    doi = {10.1162/COLI.a.16},
    url = {https://doi.org/10.1162/COLI.a.16},
    eprint = {https://direct.mit.edu/coli/article-pdf/doi/10.1162/COLI.a.16/2535477/coli.a.16.pdf},
}

@article{10.1145/3703155,
author = {Huang, Lei and Yu, Weijiang and Ma, Weitao and Zhong, Weihong and Feng, Zhangyin and Wang, Haotian and Chen, Qianglong and Peng, Weihua and Feng, Xiaocheng and Qin, Bing and Liu, Ting},
title = {A Survey on Hallucination in Large Language Models: Principles, Taxonomy, Challenges, and Open Questions},
year = {2025},
issue_date = {March 2025},
publisher = {Association for Computing Machinery},
address = {New York, NY, USA},
volume = {43},
number = {2},
issn = {1046-8188},
url = {https://doi.org/10.1145/3703155},
doi = {10.1145/3703155},
journal = {ACM Trans. Inf. Syst.},
month = jan,
articleno = {42},
numpages = {55}
}

@inproceedings{sriramananLLMCheckInvestigatingDetection2024a,
  title = {{{LLM-Check}}: {{Investigating Detection}} of {{Hallucinations}} in {{Large Language Models}}},
  booktitle = {Advances in {{Neural Information Processing Systems}}},
  author = {Sriramanan, Gaurang and Bharti, Siddhant and Sadasivan, Vinu Sankar and Saha, Shoumik and Kattakinda, Priyatham and Feizi, Soheil},
  editor = {Globerson, A. and Mackey, L. and Belgrave, D. and Fan, A. and Paquet, U. and Tomczak, J. and Zhang, C.},
  year = {2024},
  volume = {37},
  pages = {34188--34216},
  publisher = {Curran Associates, Inc.},
}

@misc{qwen2025qwen25technicalreport,
      title={Qwen2.5 Technical Report}, 
      author={Qwen and : and An Yang and Baosong Yang and Beichen Zhang and Binyuan Hui and Bo Zheng and Bowen Yu and Chengyuan Li and Dayiheng Liu and Fei Huang and Haoran Wei and Huan Lin and Jian Yang and Jianhong Tu and Jianwei Zhang and Jianxin Yang and Jiaxi Yang and Jingren Zhou and Junyang Lin and Kai Dang and Keming Lu and Keqin Bao and Kexin Yang and Le Yu and Mei Li and Mingfeng Xue and Pei Zhang and Qin Zhu and Rui Men and Runji Lin and Tianhao Li and Tianyi Tang and Tingyu Xia and Xingzhang Ren and Xuancheng Ren and Yang Fan and Yang Su and Yichang Zhang and Yu Wan and Yuqiong Liu and Zeyu Cui and Zhenru Zhang and Zihan Qiu},
      year={2025},
      eprint={2412.15115},
      archivePrefix={arXiv},
      primaryClass={cs.CL},
      url={https://arxiv.org/abs/2412.15115}, 
}

@article{radford2019language,
  title={Language Models are Unsupervised Multitask Learners},
  author={Radford, Alec and Wu, Jeff and Child, Rewon and Luan, David and Amodei, Dario and Sutskever, Ilya},
  journal={arXiv},
  year={2019}
}

@inproceedings{
renOutofDistributionDetectionSelective2022,
title={Out-of-Distribution Detection and Selective Generation for Conditional Language Models},
author={Jie Ren and Jiaming Luo and Yao Zhao and Kundan Krishna and Mohammad Saleh and Balaji Lakshminarayanan and Peter J Liu},
booktitle={The Eleventh International Conference on Learning Representations },
year={2023},
url={https://openreview.net/forum?id=kJUS5nD0vPB}
}

@misc{malininUncertaintyEstimationAutoregressive2020,
      title={Uncertainty Estimation in Autoregressive Structured Prediction}, 
      author={Andrey Malinin and Mark Gales},
      year={2021},
      eprint={2002.07650},
      archivePrefix={arXiv},
      primaryClass={stat.ML},
      url={https://arxiv.org/abs/2002.07650}, 
}

@inproceedings{manakul-etal-2023-selfcheckgpt,
    title = "{S}elf{C}heck{GPT}: Zero-Resource Black-Box Hallucination Detection for Generative Large Language Models",
    author = "Manakul, Potsawee  and
      Liusie, Adian  and
      Gales, Mark",
    editor = "Bouamor, Houda  and
      Pino, Juan  and
      Bali, Kalika",
    booktitle = "Proceedings of the 2023 Conference on Empirical Methods in Natural Language Processing",
    month = dec,
    year = "2023",
    address = "Singapore",
    publisher = "Association for Computational Linguistics",
    url = "https://aclanthology.org/2023.emnlp-main.557/",
    doi = "10.18653/v1/2023.emnlp-main.557",
    pages = "9004--9017"
}

@inproceedings{
sun2025redeep,
title={ReDe{EP}: Detecting Hallucination in Retrieval-Augmented Generation via Mechanistic Interpretability},
author={ZhongXiang Sun and Xiaoxue Zang and Kai Zheng and Jun Xu and Xiao Zhang and Weijie Yu and Yang Song and Han Li},
booktitle={The Thirteenth International Conference on Learning Representations},
year={2025},
url={https://openreview.net/forum?id=ztzZDzgfrh}
}

@misc{touvron2023llama2openfoundation,
      title={Llama 2: Open Foundation and Fine-Tuned Chat Models}, 
      author={Hugo Touvron and Louis Martin and Kevin Stone and Peter Albert and Amjad Almahairi and Yasmine Babaei and Nikolay Bashlykov and Soumya Batra and Prajjwal Bhargava and Shruti Bhosale and Dan Bikel and Lukas Blecher and Cristian Canton Ferrer and Moya Chen and Guillem Cucurull and David Esiobu and Jude Fernandes and Jeremy Fu and Wenyin Fu and Brian Fuller and Cynthia Gao and Vedanuj Goswami and Naman Goyal and Anthony Hartshorn and Saghar Hosseini and Rui Hou and Hakan Inan and Marcin Kardas and Viktor Kerkez and Madian Khabsa and Isabel Kloumann and Artem Korenev and Punit Singh Koura and Marie-Anne Lachaux and Thibaut Lavril and Jenya Lee and Diana Liskovich and Yinghai Lu and Yuning Mao and Xavier Martinet and Todor Mihaylov and Pushkar Mishra and Igor Molybog and Yixin Nie and Andrew Poulton and Jeremy Reizenstein and Rashi Rungta and Kalyan Saladi and Alan Schelten and Ruan Silva and Eric Michael Smith and Ranjan Subramanian and Xiaoqing Ellen Tan and Binh Tang and Ross Taylor and Adina Williams and Jian Xiang Kuan and Puxin Xu and Zheng Yan and Iliyan Zarov and Yuchen Zhang and Angela Fan and Melanie Kambadur and Sharan Narang and Aurelien Rodriguez and Robert Stojnic and Sergey Edunov and Thomas Scialom},
      year={2023},
      eprint={2307.09288},
      archivePrefix={arXiv},
      primaryClass={cs.CL},
      url={https://arxiv.org/abs/2307.09288}, 
}

@misc{zheng2023judging,
      title={Judging LLM-as-a-judge with MT-Bench and Chatbot Arena},
      author={Lianmin Zheng and Wei-Lin Chiang and Ying Sheng and Siyuan Zhuang and Zhanghao Wu and Yonghao Zhuang and Zi Lin and Zhuohan Li and Dacheng Li and Eric. P Xing and Hao Zhang and Joseph E. Gonzalez and Ion Stoica},
      year={2023},
      eprint={2306.05685},
      archivePrefix={arXiv},
      primaryClass={cs.CL}
}

@inproceedings{zheng2024llamafactory,
  title={LlamaFactory: Unified Efficient Fine-Tuning of 100+ Language Models},
  author={Yaowei Zheng and Richong Zhang and Junhao Zhang and Yanhan Ye and Zheyan Luo and Zhangchi Feng and Yongqiang Ma},
  booktitle={Proceedings of the 62nd Annual Meeting of the Association for Computational Linguistics (Volume 3: System Demonstrations)},
  address={Bangkok, Thailand},
  publisher={Association for Computational Linguistics},
  year={2024},
  url={http://arxiv.org/abs/2403.13372}
}

@misc{alpaca,
  author = {Rohan Taori and Ishaan Gulrajani and Tianyi Zhang and Yann Dubois and Xuechen Li and Carlos Guestrin and Percy Liang and Tatsunori B. Hashimoto },
  title = {Stanford Alpaca: An Instruction-following LLaMA model},
  year = {2023},
  publisher = {GitHub},
  journal = {GitHub repository},
  howpublished = {\url{https://github.com/tatsu-lab/stanford_alpaca}},
}

@misc{grattafiori2024llama3herdmodels,
      title={The Llama 3 Herd of Models}, 
      author={Aaron Grattafiori and Abhimanyu Dubey and Abhinav Jauhri and Abhinav Pandey and Abhishek Kadian and Ahmad Al-Dahle and Aiesha Letman and Akhil Mathur and Alan Schelten and Alex Vaughan and Amy Yang and Angela Fan and Anirudh Goyal and Anthony Hartshorn and Aobo Yang and Archi Mitra and Archie Sravankumar and Artem Korenev and Arthur Hinsvark and Arun Rao and Aston Zhang and Aurelien Rodriguez and Austen Gregerson and Ava Spataru and Baptiste Roziere and Bethany Biron and Binh Tang and Bobbie Chern and Charlotte Caucheteux and Chaya Nayak and Chloe Bi and Chris Marra and Chris McConnell and Christian Keller and Christophe Touret and Chunyang Wu and Corinne Wong and Cristian Canton Ferrer and Cyrus Nikolaidis and Damien Allonsius and Daniel Song and Danielle Pintz and Danny Livshits and Danny Wyatt and David Esiobu and Dhruv Choudhary and Dhruv Mahajan and Diego Garcia-Olano and Diego Perino and Dieuwke Hupkes and Egor Lakomkin and Ehab AlBadawy and Elina Lobanova and Emily Dinan and Eric Michael Smith and Filip Radenovic and Francisco Guzmán and Frank Zhang and Gabriel Synnaeve and Gabrielle Lee and Georgia Lewis Anderson and Govind Thattai and Graeme Nail and Gregoire Mialon and Guan Pang and Guillem Cucurell and Hailey Nguyen and Hannah Korevaar and Hu Xu and Hugo Touvron and Iliyan Zarov and Imanol Arrieta Ibarra and Isabel Kloumann and Ishan Misra and Ivan Evtimov and Jack Zhang and Jade Copet and Jaewon Lee and Jan Geffert and Jana Vranes and Jason Park and Jay Mahadeokar and Jeet Shah and Jelmer van der Linde and Jennifer Billock and Jenny Hong and Jenya Lee and Jeremy Fu and Jianfeng Chi and Jianyu Huang and Jiawen Liu and Jie Wang and Jiecao Yu and Joanna Bitton and Joe Spisak and Jongsoo Park and Joseph Rocca and Joshua Johnstun and Joshua Saxe and Junteng Jia and Kalyan Vasuden Alwala and Karthik Prasad and Kartikeya Upasani and Kate Plawiak and Ke Li and Kenneth Heafield and Kevin Stone and Khalid El-Arini and Krithika Iyer and Kshitiz Malik and Kuenley Chiu and Kunal Bhalla and Kushal Lakhotia and Lauren Rantala-Yeary and Laurens van der Maaten and Lawrence Chen and Liang Tan and Liz Jenkins and Louis Martin and Lovish Madaan and Lubo Malo and Lukas Blecher and Lukas Landzaat and Luke de Oliveira and Madeline Muzzi and Mahesh Pasupuleti and Mannat Singh and Manohar Paluri and Marcin Kardas and Maria Tsimpoukelli and Mathew Oldham and Mathieu Rita and Maya Pavlova and Melanie Kambadur and Mike Lewis and Min Si and Mitesh Kumar Singh and Mona Hassan and Naman Goyal and Narjes Torabi and Nikolay Bashlykov and Nikolay Bogoychev and Niladri Chatterji and Ning Zhang and Olivier Duchenne and Onur Çelebi and Patrick Alrassy and Pengchuan Zhang and Pengwei Li and Petar Vasic and Peter Weng and Prajjwal Bhargava and Pratik Dubal and Praveen Krishnan and Punit Singh Koura and Puxin Xu and Qing He and Qingxiao Dong and Ragavan Srinivasan and Raj Ganapathy and Ramon Calderer and Ricardo Silveira Cabral and Robert Stojnic and Roberta Raileanu and Rohan Maheswari and Rohit Girdhar and Rohit Patel and Romain Sauvestre and Ronnie Polidoro and Roshan Sumbaly and Ross Taylor and Ruan Silva and Rui Hou and Rui Wang and Saghar Hosseini and Sahana Chennabasappa and Sanjay Singh and Sean Bell and Seohyun Sonia Kim and Sergey Edunov and Shaoliang Nie and Sharan Narang and Sharath Raparthy and Sheng Shen and Shengye Wan and Shruti Bhosale and Shun Zhang and Simon Vandenhende and Soumya Batra and Spencer Whitman and Sten Sootla and Stephane Collot and Suchin Gururangan and Sydney Borodinsky and Tamar Herman and Tara Fowler and Tarek Sheasha and Thomas Georgiou and Thomas Scialom and Tobias Speckbacher and Todor Mihaylov and Tong Xiao and Ujjwal Karn and Vedanuj Goswami and Vibhor Gupta and Vignesh Ramanathan and Viktor Kerkez and Vincent Gonguet and Virginie Do and Vish Vogeti and Vítor Albiero and Vladan Petrovic and Weiwei Chu and Wenhan Xiong and Wenyin Fu and Whitney Meers and Xavier Martinet and Xiaodong Wang and Xiaofang Wang and Xiaoqing Ellen Tan and Xide Xia and Xinfeng Xie and Xuchao Jia and Xuewei Wang and Yaelle Goldschlag and Yashesh Gaur and Yasmine Babaei and Yi Wen and Yiwen Song and Yuchen Zhang and Yue Li and Yuning Mao and Zacharie Delpierre Coudert and Zheng Yan and Zhengxing Chen and Zoe Papakipos and Aaditya Singh and Aayushi Srivastava and Abha Jain and Adam Kelsey and Adam Shajnfeld and Adithya Gangidi and Adolfo Victoria and Ahuva Goldstand and Ajay Menon and Ajay Sharma and Alex Boesenberg and Alexei Baevski and Allie Feinstein and Amanda Kallet and Amit Sangani and Amos Teo and Anam Yunus and Andrei Lupu and Andres Alvarado and Andrew Caples and Andrew Gu and Andrew Ho and Andrew Poulton and Andrew Ryan and Ankit Ramchandani and Annie Dong and Annie Franco and Anuj Goyal and Aparajita Saraf and Arkabandhu Chowdhury and Ashley Gabriel and Ashwin Bharambe and Assaf Eisenman and Azadeh Yazdan and Beau James and Ben Maurer and Benjamin Leonhardi and Bernie Huang and Beth Loyd and Beto De Paola and Bhargavi Paranjape and Bing Liu and Bo Wu and Boyu Ni and Braden Hancock and Bram Wasti and Brandon Spence and Brani Stojkovic and Brian Gamido and Britt Montalvo and Carl Parker and Carly Burton and Catalina Mejia and Ce Liu and Changhan Wang and Changkyu Kim and Chao Zhou and Chester Hu and Ching-Hsiang Chu and Chris Cai and Chris Tindal and Christoph Feichtenhofer and Cynthia Gao and Damon Civin and Dana Beaty and Daniel Kreymer and Daniel Li and David Adkins and David Xu and Davide Testuggine and Delia David and Devi Parikh and Diana Liskovich and Didem Foss and Dingkang Wang and Duc Le and Dustin Holland and Edward Dowling and Eissa Jamil and Elaine Montgomery and Eleonora Presani and Emily Hahn and Emily Wood and Eric-Tuan Le and Erik Brinkman and Esteban Arcaute and Evan Dunbar and Evan Smothers and Fei Sun and Felix Kreuk and Feng Tian and Filippos Kokkinos and Firat Ozgenel and Francesco Caggioni and Frank Kanayet and Frank Seide and Gabriela Medina Florez and Gabriella Schwarz and Gada Badeer and Georgia Swee and Gil Halpern and Grant Herman and Grigory Sizov and Guangyi and Zhang and Guna Lakshminarayanan and Hakan Inan and Hamid Shojanazeri and Han Zou and Hannah Wang and Hanwen Zha and Haroun Habeeb and Harrison Rudolph and Helen Suk and Henry Aspegren and Hunter Goldman and Hongyuan Zhan and Ibrahim Damlaj and Igor Molybog and Igor Tufanov and Ilias Leontiadis and Irina-Elena Veliche and Itai Gat and Jake Weissman and James Geboski and James Kohli and Janice Lam and Japhet Asher and Jean-Baptiste Gaya and Jeff Marcus and Jeff Tang and Jennifer Chan and Jenny Zhen and Jeremy Reizenstein and Jeremy Teboul and Jessica Zhong and Jian Jin and Jingyi Yang and Joe Cummings and Jon Carvill and Jon Shepard and Jonathan McPhie and Jonathan Torres and Josh Ginsburg and Junjie Wang and Kai Wu and Kam Hou U and Karan Saxena and Kartikay Khandelwal and Katayoun Zand and Kathy Matosich and Kaushik Veeraraghavan and Kelly Michelena and Keqian Li and Kiran Jagadeesh and Kun Huang and Kunal Chawla and Kyle Huang and Lailin Chen and Lakshya Garg and Lavender A and Leandro Silva and Lee Bell and Lei Zhang and Liangpeng Guo and Licheng Yu and Liron Moshkovich and Luca Wehrstedt and Madian Khabsa and Manav Avalani and Manish Bhatt and Martynas Mankus and Matan Hasson and Matthew Lennie and Matthias Reso and Maxim Groshev and Maxim Naumov and Maya Lathi and Meghan Keneally and Miao Liu and Michael L. Seltzer and Michal Valko and Michelle Restrepo and Mihir Patel and Mik Vyatskov and Mikayel Samvelyan and Mike Clark and Mike Macey and Mike Wang and Miquel Jubert Hermoso and Mo Metanat and Mohammad Rastegari and Munish Bansal and Nandhini Santhanam and Natascha Parks and Natasha White and Navyata Bawa and Nayan Singhal and Nick Egebo and Nicolas Usunier and Nikhil Mehta and Nikolay Pavlovich Laptev and Ning Dong and Norman Cheng and Oleg Chernoguz and Olivia Hart and Omkar Salpekar and Ozlem Kalinli and Parkin Kent and Parth Parekh and Paul Saab and Pavan Balaji and Pedro Rittner and Philip Bontrager and Pierre Roux and Piotr Dollar and Polina Zvyagina and Prashant Ratanchandani and Pritish Yuvraj and Qian Liang and Rachad Alao and Rachel Rodriguez and Rafi Ayub and Raghotham Murthy and Raghu Nayani and Rahul Mitra and Rangaprabhu Parthasarathy and Raymond Li and Rebekkah Hogan and Robin Battey and Rocky Wang and Russ Howes and Ruty Rinott and Sachin Mehta and Sachin Siby and Sai Jayesh Bondu and Samyak Datta and Sara Chugh and Sara Hunt and Sargun Dhillon and Sasha Sidorov and Satadru Pan and Saurabh Mahajan and Saurabh Verma and Seiji Yamamoto and Sharadh Ramaswamy and Shaun Lindsay and Shaun Lindsay and Sheng Feng and Shenghao Lin and Shengxin Cindy Zha and Shishir Patil and Shiva Shankar and Shuqiang Zhang and Shuqiang Zhang and Sinong Wang and Sneha Agarwal and Soji Sajuyigbe and Soumith Chintala and Stephanie Max and Stephen Chen and Steve Kehoe and Steve Satterfield and Sudarshan Govindaprasad and Sumit Gupta and Summer Deng and Sungmin Cho and Sunny Virk and Suraj Subramanian and Sy Choudhury and Sydney Goldman and Tal Remez and Tamar Glaser and Tamara Best and Thilo Koehler and Thomas Robinson and Tianhe Li and Tianjun Zhang and Tim Matthews and Timothy Chou and Tzook Shaked and Varun Vontimitta and Victoria Ajayi and Victoria Montanez and Vijai Mohan and Vinay Satish Kumar and Vishal Mangla and Vlad Ionescu and Vlad Poenaru and Vlad Tiberiu Mihailescu and Vladimir Ivanov and Wei Li and Wenchen Wang and Wenwen Jiang and Wes Bouaziz and Will Constable and Xiaocheng Tang and Xiaojian Wu and Xiaolan Wang and Xilun Wu and Xinbo Gao and Yaniv Kleinman and Yanjun Chen and Ye Hu and Ye Jia and Ye Qi and Yenda Li and Yilin Zhang and Ying Zhang and Yossi Adi and Youngjin Nam and Yu and Wang and Yu Zhao and Yuchen Hao and Yundi Qian and Yunlu Li and Yuzi He and Zach Rait and Zachary DeVito and Zef Rosnbrick and Zhaoduo Wen and Zhenyu Yang and Zhiwei Zhao and Zhiyu Ma},
      year={2024},
      eprint={2407.21783},
      archivePrefix={arXiv},
      primaryClass={cs.AI},
      url={https://arxiv.org/abs/2407.21783}, 
}

@article{kuhn2023semantic,
  title={Semantic uncertainty: Linguistic invariances for uncertainty estimation in natural language generation},
  author={Kuhn, Lorenz and Gal, Yarin and Farquhar, Sebastian},
  journal={arXiv preprint arXiv:2302.09664},
  year={2023}
}

@inproceedings{zha2023alignscore,
    title = "{A}lign{S}core: Evaluating Factual Consistency with A Unified Alignment Function",
    author = "Zha, Yuheng  and
      Yang, Yichi  and
      Li, Ruichen  and
      Hu, Zhiting",
    editor = "Rogers, Anna  and
      Boyd-Graber, Jordan  and
      Okazaki, Naoaki",
    booktitle = "Proceedings of the 61st Annual Meeting of the Association for Computational Linguistics (Volume 1: Long Papers)",
    month = jul,
    year = "2023",
    address = "Toronto, Canada",
    publisher = "Association for Computational Linguistics",
    url = "https://aclanthology.org/2023.acl-long.634/",
    doi = "10.18653/v1/2023.acl-long.634",
    pages = "11328--11348"
}

@inproceedings{zhang2025icrprobetrackinghidden,
    title = "{ICR} Probe: Tracking Hidden State Dynamics for Reliable Hallucination Detection in {LLM}s",
    author = "Zhang, Zhenliang  and
      Hu, Xinyu  and
      Zhang, Huixuan  and
      Zhang, Junzhe  and
      Wan, Xiaojun",
    editor = "Che, Wanxiang  and
      Nabende, Joyce  and
      Shutova, Ekaterina  and
      Pilehvar, Mohammad Taher",
    booktitle = "Proceedings of the 63rd Annual Meeting of the Association for Computational Linguistics (Volume 1: Long Papers)",
    month = jul,
    year = "2025",
    address = "Vienna, Austria",
    publisher = "Association for Computational Linguistics",
    url = "https://aclanthology.org/2025.acl-long.880/",
    doi = "10.18653/v1/2025.acl-long.880",
    pages = "17986--18002",
    ISBN = "979-8-89176-251-0"
}

\appendix

\section{Appendix}
\label{sec:appendix}

\subsection{Acknowledgment of LLM Usage}
We used AI-assisted tools (e.g., ChatGPT) for light copyediting (grammar, word choice, and clarity) in portions of the paper. We also used it to assist in checking the recency of citations during the literature review by surfacing potentially relevant recent work. All suggestions were reviewed and verified by the authors; the study design, analyses, claims, and final text are our own.

\subsection{Baseline Methods}
\label{appendix:baselines}

\paragraph{PPL (Perplexity)}
Perplexity measures how well the model predicts its own generated answer by aggregating the next-token negative log-likelihood under teacher forcing. For an answer sequence $y_{1:T}$, we define
$\mathrm{PPL} = \exp\!\Big(-\tfrac{1}{T}\sum_{t=1}^{T}\log p_\theta(y_t \mid y_{<t}, x)\Big)$.
Larger $\mathrm{PPL}$ indicates lower token-level confidence and is often associated with less reliable (and potentially hallucinated) outputs. Because it is computed from a single forward pass on the model’s own answer, $\mathrm{PPL}$ is supervision-free and incurs no additional decoding cost.

The teacher-forced log-probability for token position $t$ is gathered from the \emph{previous} timestep’s logits (standard LM shift): $\log p_\theta(y_t \mid y_{<t}, x) = \log \mathrm{softmax}(L_{t-1})[y_t]$. When scoring an answer window that starts at token index $i_1$ and ends at $i_2$, we therefore use logits on indices $[i_1{-}1,\,i_2{-}1]$ and target tokens on $[i_1,\,i_2]$; this requires $i_1>0$.

\paragraph{Window-Entropy}
We use an entropy-based uncertainty score over an answer window under teacher forcing. Let $H_t = H\!\big(p_\theta(\cdot \mid y_{<t}, x)\big)$ denote the categorical entropy of the next-token distribution at position $t$. We operationalize Window-Entropy as a windowed entropy statistic:
\[
\mathrm{Win}\text{-}\mathrm{Ent}_w \;=\; \max_{\text{$w$-window}} \frac{1}{w}\sum_{t\in \text{window}} H_t,
\]
and use $w{=}1$, which reduces to the maximum token entropy over the answer window.

\paragraph{Semantic Entropy Probes (SEP)} SEP avoids multi-sample decoding by training a lightweight probe to predict a semantic-uncertainty score (e.g., SE) directly from the model’s internal representations. Concretely, SEP fits a small head (often linear) on top of a selected hidden state (e.g., a fixed position near answer onset) using SE-derived targets. At inference time, SEP is single-pass: extract the hidden state from one forward pass and apply the probe to obtain the score, yielding near-SE discriminative behavior at a cost comparable to other telemetry-based metrics. Because SEP relies on SE targets during training, it is supervision-based even though its test-time computation is cheap.

\paragraph{LLM-Check}
LLM-Check is a single-pass hallucination detector that fuses token-level uncertainty with representation-based signals extracted from hidden states and attentions under teacher forcing. Specifically, it computes geometric features summarizing the internal trajectory, including mean log-determinant measures of per-layer hidden-state covariance (\textit{Hidden Score}) and attention-kernel maps (\textit{Attention Score}). At inference time LLM-Check is training-free and often improves over uncertainty-only baselines across diverse evaluation settings.

PPL and Window-Entropy are single-pass and training-free. SEP is a training-based but single-pass at inference. LLM-Check is single-pass and training-free (feature computation only).

\paragraph{ICR-probe}
ICR-probe characterizes \emph{cross-layer residual-stream dynamics} by comparing them to attention-based routing.
For token $i$ at layer $\ell$, let $x_j^\ell$ denote the hidden state of token $j$, let $X^\ell \in \mathbb{R}^{T\times D}$ be the layer hidden-state matrix, and define the residual update as $\Delta x_i^\ell = x_i^\ell - x_i^{\ell-1}$.
We induce a distribution over \emph{source} tokens by scoring how well the update aligns with token directions:
\[
s_{ij}^\ell \propto \left\langle \Delta x_i^\ell,\; x_j^\ell \right\rangle
\quad\Rightarrow\quad
\textstyle \mathrm{Proj}_i^\ell(j) = \frac{\exp(s_{ij}^\ell)}{\sum_{k}\exp(s_{ik}^\ell)}.
\]
Let $\mathrm{Attn}_i^\ell$ be the head-averaged attention distribution for token $i$ at layer $\ell$ (computed under teacher forcing on the generated answer).
ICR then measures the mismatch between where the residual stream is \emph{actually updated} and where attention \emph{allocates} contextual influence: $\mathrm{ICR}_i^\ell \;=\; \mathrm{JSD}\!\big(\mathrm{Proj}_i^\ell,\; \mathrm{Attn}_i^\ell\big)$. To reduce noise and emphasize salient context, we restrict both distributions to the top-$k$ tokens by attention weight (e.g., $k{=}20$), and renormalize after truncation.

\paragraph{Pooling and feature construction}
We compute ICR over answer tokens within an \emph{answer window} $W$ defined by the prompt boundary.
For each layer $\ell$, we average token-wise scores across $i\in W$ to obtain a layer-level feature: $\mathbf{v} \in \mathbb{R}^{L},\space v_\ell = \tfrac{1}{|W|}\sum_{i\in W}\mathrm{ICR}_i^\ell.$ Larger values of $v_\ell$ indicate greater divergence between residual updates and attention-based routing, consistent with increased ``parametric injection'' that is not accounted for by attention-mediated context reuse. We use a small MLP to map the layer-wise feature vector to a sequence-level hallucination probability.

\begin{table*}[t]
\centering
\small
\setlength{\tabcolsep}{5pt}
\begin{tabular}{@{}l c c c c c c c c@{}}
\toprule
& \textbf{Baseline} & \textbf{All} & \textbf{HV+AD} & \textbf{TE+AD} & \textbf{TE+HV} & \textbf{TE only} & \textbf{HV only} & \textbf{AD only} \\
\midrule
PPL (AUROC)        & 53.18 & $\Delta$-5.10 & $\Delta$-3.69 & $\Delta$-0.61 & $\Delta$-4.10 & $\Delta$-1.74 & $\Delta$-4.58 & $\Delta$+0.34 \\
PPL (ACC)          & 58.68 & $\Delta$-7.78 & $\Delta$-8.08 & $\Delta$-8.68 & $\Delta$-7.78 & $\Delta$-8.68 & $\Delta$-8.08 & $\Delta$-8.98 \\

\hline
Window-Entropy (AUROC)     & 56.99 & $\Delta$-1.69 & $\Delta$-5.35 & $\Delta$+7.24 & $\Delta$-4.20 & $\Delta$+0.95 & $\Delta$+4.61 & $\Delta$+3.10 \\
Window-Entropy (ACC)       & 57.19 & $\Delta$-8.69 & $\Delta$-7.78 & $\Delta$-7.19 & $\Delta$-2.40 & $\Delta$-6.89 & $\Delta$-0.90 & $\Delta$-7.19 \\

\hline
Hidden L17 (AUROC) & 58.15 & $\Delta$-4.04 & $\Delta$-1.67 & $\Delta$+0.39 & $\Delta$-2.48 & $\Delta$+0.65 & $\Delta$-2.38 & $\Delta$+0.37 \\
Hidden L17 (ACC)   & 56.89 & $\Delta$-6.89 & $\Delta$-6.89 & $\Delta$-2.69 & $\Delta$-6.89 & $\Delta$-1.80 & $\Delta$-6.89 & $\Delta$-2.99 \\

\hline
Attn L20 (AUROC)   & 73.63 & $\Delta$-25.54 & $\Delta$-19.63 & $\Delta$-19.11 & $\Delta$-22.80 & $\Delta$-2.59 & $\Delta$-22.27 & $ \Delta$-20.97 \\
Attn L20 (ACC)     & 69.76 & $\Delta$-19.76 & $\Delta$-19.76 & $\Delta$-19.76 & $\Delta$-19.76 & $\Delta$-5.39 & $\Delta$-19.76 & $\Delta$-19.76 \\

\bottomrule
\end{tabular} 
\caption{\textbf{Training-free single-pass detector scores under term ablations on FAVA-Annotation (Llama-2).} The first column shows the original baseline scores; subsequent columns show deltas ($\Delta$ = adapted $-$ clean) for CORVUS variants trained with the indicated telemetry terms enabled in Eq.~\ref{eq:methods-obj} (others set to zero; $\mathcal{L}_{\mathrm{CE}}$ always included). AUROC and ACC are measured at the clean-calibrated threshold.}
\label{tab:ablation-terms}
\end{table*}

\paragraph{Notes and practicalities}
\begin{itemize}
  \item \textbf{Top-$k$ restriction.} Truncating both distributions to the top-$k$ tokens by attention weight suppresses long-tail noise and improves numerical stability.
  \item \textbf{Answer window.} Aggregating over an identified answer span reduces prompt-induced artifacts and targets model-generated content.
  \item \textbf{Access requirements.} ICR-probe assumes white-box (or instrumented) access to hidden states and attention matrices; it does not rely on additional sampling or external retrieval.
\end{itemize}

\begin{table*}[t]
\centering
\small
\setlength{\tabcolsep}{7pt}
\begin{tabular}{@{}l c c c c c@{}}
\toprule
& \textbf{Baseline} & \textbf{No FGSM} & \textbf{$\varepsilon{=}1.5\!\times\!10^{-3}$} & \textbf{$\varepsilon{=}5\!\times\!10^{-3}$} & \textbf{($\varepsilon{=}1.0\!\times\!10^{-2}$, CORVUS)} \\
\midrule
PPL (AUROC)        & 53.18 & $\Delta$-4.10 & $\Delta$-1.73 & $\Delta$-3.85 & $\Delta$-5.10 \\
PPL (ACC)          & 58.68 & $\Delta$-7.78 & $\Delta$-8.38  & $\Delta$-7.78 & $\Delta$-7.78 \\
\hline
Window-Entropy (AUROC)     & 56.99 & $\Delta$-4.20 & $\Delta$+1.18 & $\Delta$+2.25 & $\Delta$-1.69 \\
Window-Entropy (ACC)       & 57.19 & $\Delta$-2.40 & $\Delta$-7.49 & $\Delta$-7.78 & $\Delta$-8.69 \\
\hline
Hidden L17 (AUROC) & 58.15 & $\Delta$-2.48 & $\Delta$-2.46 & $\Delta$-3.18 & $\Delta$-4.04 \\
Hidden L17 (ACC)   & 56.89 & $\Delta$-6.89 & $\Delta$-6.89 & $\Delta$-6.89 & $\Delta$-6.89 \\
\hline
Attn L20 (AUROC)   & 73.63 & $\Delta$-14.80 & $\Delta$-16.80 & $\Delta$-19.50 & $\Delta$-25.54 \\
Attn L20 (ACC)     & 69.76 & $\Delta$-19.76 & $\Delta$-19.76 & $\Delta$-19.76 & $\Delta$-19.76 \\
\bottomrule
\end{tabular}
\caption{\textbf{FGSM sweep on FAVA-Annotation (Baseline).} Effect of the embedding-step size $\varepsilon$ on $\Delta$AUROC/$\Delta$ACC (attack $-$ clean) for an LLM-Check detector evaluated on paraphrased prompts. Columns report deltas at $\varepsilon{=}0$ (no FGSM), $1.5\!\times\!10^{-3}$, $5\!\times\!10^{-3}$, and  the default $1.0\!\times\!10^{-2}$ of CORVUS.}
\label{tab:ablation-fgsm}
\end{table*}

\section{Ablation Studies}
\label{sec:ablations}

We next study which parts of the CORVUS objective contribute most to evasion. We focus on representative training-free single-pass methods (i.e., PPL, Window-Entropy, and LLM-Check components) because they cover the telemetry families targeted by CORVUS. Unless otherwise noted, we report $\Delta$AUROC (attack minus clean, so more negative values indicate stronger evasion) on \emph{FAVA-Annotation}. A representative setting uses Llama-2-7B-Chat with LoRA rank $r{=}64$, scaling $\alpha{=}128$, attention/hidden layer band 50--75\% of layers, $\lambda_{\mathrm{adv}}{=}0.5$, $(\beta_{\mathrm{te}},\beta_{\mathrm{hv}}){=}(0.07,0.05)$, and FGSM step size $\varepsilon{=}1.0{\times}10^{-2}$.

\subsection{Per-term Ablation}
\label{sec:ablations:terms}

We ablate the telemetry-shaping terms in the CORVUS objective to identify which signals are responsible for evading LLM-Check.
Concretely, we train CORVUS variants that keep only selected subsets of
$\{\mathrm{TE}_{\mathrm{clean}}, \overline{\mathrm{HV}}_{\mathrm{clean}}, \overline{\mathrm{AD}}_{\mathrm{adv}}\}$
from Eq.~\ref{eq:methods-obj}, while holding all other training details fixed. Column headers in Table~\ref{tab:ablation-terms} indicate which of these telemetry-shaping terms are enabled (others set to zero; $\mathcal{L}_{\mathrm{CE}}$ is always included). Thus, \emph{AD only} keeps only $\overline{\mathrm{AD}}_{\mathrm{adv}}$, while \emph{TE+HV} corresponds to removing the AD term. We then evaluate LLM-Check on FAVA-Annotation and report the change in AUROC. Recall that $\overline{\mathrm{AD}}_{\mathrm{adv}}$ is applied only on the adversarial (FGSM) pass and encourages higher attention diagonality.

Table~\ref{tab:ablation-terms} shows a clear separation of roles across terms. First, the attention-diagonality term is the most effective single contributor for degrading the attention-based component of LLM-Check: using $\overline{\mathrm{AD}}_{\mathrm{adv}}$ alone produces the largest AUROC drop for the attention score, and adding it to other terms generally strengthens attention-score evasion. Second, the hidden log-volume term exhibits more mixed behavior under an AUROC lens: it can meaningfully shift hidden-state telemetry (and can worsen fixed operating points), but its effect on ranking performance is not uniformly dominant across components. Third, token-entropy suppression by itself provides only a small change in evasion, and its effect is detector-component dependent; empirically, it behaves more like a complementary knob than a primary mechanism.

Overall, these results suggest that CORVUS does not rely on a single monolithic failure mode of LLM-Check. Instead, $\overline{\mathrm{AD}}_{\mathrm{adv}}$ is the main lever for evading attention-centric features, while $\mathrm{TE}_{\mathrm{clean}}$ and $\overline{\mathrm{HV}}_{\mathrm{clean}}$ provide secondary shaping that can further perturb uncertainty- and representation-based telemetry depending on the component and operating point.

\subsection{Role of the AD-targeted perturbation}
\label{sec:ablations:fgsm}

This ablation isolates the effect of the AD-targeted FGSM pass. The adversarial perturbation is designed to maximize the surrogate detector score $-\mathrm{AD}$ (equivalently, to minimize attention diagonality), producing a proxy pass whose resulting $AD_{\mathrm{adv}}$ supplies an additional attention-shaping signal during clean-only training. To quantify its contribution, the embedding-space step size $\varepsilon$ is varied and detectors are evaluated on FAVA-Annotation, including paraphrased prompts as a stress test.

As shown in Table~\ref{tab:ablation-fgsm}, removing the adversarial pass ($\varepsilon{=}0$) consistently reduces evasion under paraphrases, indicating that the AD-targeted proxy pass strengthens the learned attention-telemetry shift beyond what the clean pass alone provides. For nonzero steps, the most effective $\varepsilon$ depends on which telemetry family dominates a detector: smaller-to-moderate steps (e.g., $\varepsilon \approx 1.5{\times}10^{-3}$) most strongly degrade uncertainty/entropy-driven signals, while larger steps can further amplify degradation of attention-based telemetry. The main experiments therefore use a moderate default that yields strong paraphrase transfer without overly emphasizing a single telemetry component.

\paragraph{Attack-calibrated operating point (oracle).}
Fixed-threshold deltas (calibrated on the clean model and held fixed) reflect both (i) calibration shift and (ii) changes in ranking separability. To disentangle these, we additionally report an \emph{attack-calibrated} operating point that \emph{reselects} the threshold on attacked telemetry to achieve a target false-positive rate, providing an oracle upper bound on what an auditor could recover if re-calibration were allowed after model-side adaptation. Table~\ref{tab:attack-calibrated-eps1e-2} reports this attack-calibrated operating point for $\varepsilon{=}1.0{\times}10^{-2}$, including TPR@5\%FPR and F1. Notably, the attention score remains near chance even after re-calibration, consistent with a separability collapse rather than a pure threshold shift.

\begin{table}[t]
\centering
\small
\setlength{\tabcolsep}{4pt}
\begin{tabular}{@{}l c c c c@{}}
\toprule
\textbf{Score} & \textbf{AUROC} & \textbf{FPR} & \textbf{TPR@5\%FPR} & \textbf{F1} \\
\midrule
PPL            & 48.09 & 2.40 & 6.59  & 12.09 \\
Window-Entropy & 55.30 & 4.79 & 2.99  & 5.56 \\
Hidden (L17)   & 54.10 & 4.19 & 10.78 & 18.75 \\
Attn (L20)     & 48.08 & 2.99 & 1.80  & 3.43 \\
\bottomrule
\end{tabular}
\caption{\textbf{Attack-calibrated operating point at $\varepsilon{=}1.0{\times}10^{-2}$.} Thresholds are selected \emph{on attacked telemetry} to achieve \(\le 5\%\) FPR (discrete thresholds can yield FPR $<5\%$), then TPR@5\%FPR and F1 are reported at that threshold. Metrics use the same balanced FAVA-Annotation subset as the other tables.}
\label{tab:attack-calibrated-eps1e-2}
\end{table}

\paragraph{Takeaways.}
The ablations suggest three points:
(i) AD shaping is a primary driver of evasion for attention-centric LLM-Check features;
(ii) TE and HV provide complementary shaping signals whose impact varies by detector component and operating point; and
(iii) the AD-targeted FGSM pass acts as a telemetry-directed proxy signal that improves paraphrase transfer of the induced internal-signal shift, rather than robustness training or a requirement for in-distribution evasion.
\end{document}